\documentstyle[12pt]{article}
\def \be{\begin{equation}}
\def \ee{\end{equation}}
\def \bea{\begin{eqnarray}}
\def \eea{\end{eqnarray}}

\title{Hamiltonian Lattice QCD at Finite Chemical Potential}

\author{
Eric B. Gregory$^{1}$, ~
Shuo-Hong Guo$^{1}$,   ~
Helmut Kr\"oger$^{2}$ \thanks{E-mail: hkroger@phy.ulaval.ca},\\ 
Xiang-Qian Luo$^{3,1}$\thanks{Corresponding author. E-mail: stslxq@zsu.edu.cn}
\vspace{2cm}
\\
$^1${\small\sl Department of Physics, Zhongshan University, 
Guangzhou 510275, China
\thanks{Mailing Address}
}\\
$^2${\small\sl D\'epartement de Physique, Universit\'e Laval, Qu\'ebec, 
Qu\'ebec G1K 7P4, Canada}\\
$^3${\small\sl CCAST (World Laboratory), P.O. Box 8730, 
Beijing 100080, China}\\
}

\begin{document}
\maketitle

\begin{flushleft}
{\bf Abstract}
\end{flushleft}
\noindent
At sufficiently high temperature and density, quantum chromodynamics (QCD) is 
expected to undergo a phase transition 
from the confined phase to the quark-gluon plasma phase. 
In the Lagrangian lattice formulation 
the Monte Carlo method works well 
for QCD at finite temperature, however, 
it breaks down at finite chemical potential. 
We develop a Hamiltonian approach to lattice QCD at finite chemical 
potential and solve it in the case of free quarks and 
in the strong coupling limit. 
At zero temperature, we calculate the vacuum energy,
chiral condensate, quark number density and its susceptibility, 
as well as mass of the pseudoscalar, vector mesons and nucleon. We find that
the chiral phase transition is of first order, 
and the critical chemical potential is $\mu_C =m_{dyn}^{(0)}$ 
(dynamical quark mass at $\mu=0$). This is
consistent with $\mu_C \approx M_N^{(0)}/3$ 
(where $M_N^{(0)}$ is the nucleon mass at $\mu=0$).

\newpage

\section{Introduction}
\subsection{Motivation}
According to the big bang model in cosmology, the early universe 
underwent a series of drastic changes. 
For some time it was a hot and dense quark-gluon plasma (QGP), where quarks 
and gluons were deconfined. Today it is in a low temperature 
and low density hadronic phase, where quarks are confined.
The ultimate goal of machines such as the Relativistic Heavy Ion Collider (RHIC) 
at BNL and the Large Hadron Collider (LHC) at CERN is to create the QGP phase.  
The QGP may also exist in the core of very dense stars such as neutron stars.
Quantum chromodynamics (QCD) is the fundamental 
theory of quarks and gluons.
A precise determination of the QCD phase structure
at finite temperature $T$ and chemical potential $\mu$ will  
provide valuable information in the  experimental search for the QGP.
Lattice gauge theory (LGT) 
proposed by Wilson in 1974, is a very reliable technique 
for the investigation of phase transitions.
There are no free parameters
in LGT when the continuum limit is taken, 
in contrast with other nonperturbative techniques.
Although the standard lattice Lagrangian Monte 
Carlo method works very well 
for QCD at finite temperature,
it  unfortunately breaks down at finite chemical potential 
(due to the so-called complex action problem). This is briefly summarized in
Sect. \ref{Status}.
On the other hand, lattice QCD at finite chemical potential formulated 
in the Hamiltonian approach does not encounter a complex action problem.
In Sect. \ref{Our_Approach},
we develop a Hamiltonian approach to lattice QCD at finite chemical 
potential $\mu$. We solve this  
in the case of free quarks and in the strong coupling limit.

\subsection{Present status}
\label{Status}
LGT is an approach to QCD from first principles. 
However, it is not free of problems:
(a) First, there are lattice artifacts: 
A finite volume and a finite lattice spacing introduce errors.
(b) There is a no-go theorem for chiral fermions: 
There is species doubling of any local fermionic theory 
with continuous symmetries. 
For naive fermions, chiral symmetry is preserved,
but the species are doubled and the chiral anomaly is wrong.
Kogut-Susskind fermions preserve the continuous U(1) chiral symmetry,
but break explicitly flavor symmetry. 
For Wilson fermions, the flavor symmetry exists,
but chiral symmetry is explicitly broken.
Kogut-Susskind fermions and Wilson fermions have been extensively used
in numerical simulations. Recently, there has been evidence showing
that those two approaches may give the topological charge or anomaly
incorrectly \cite{Topological} on a finite lattice.
Therefore, it is far from clear whether correct results in the continuum 
can be obtained using those fermion formulations.
Kaplan's domain wall fermions \cite{Kaplan} 
and Neuberger's overlap fermion formulation 
\cite{Neuberger} have attracted much attention, 
because they give the correct chiral modes, they also produce the correct 
anomaly and topological charge. 
For domain wall fermions there is an extra dimension and the lattice size 
in this dimension has to be very large.
Thus algorithms suitable for those new fermion approaches 
need to be developed.
In this paper, we do not address those problems.

\bigskip

\noindent Here we would like to investigate lattice QCD 
at finite chemical potential. 
In the continuum, the grand canonical partition function of QCD 
at finite temperature $T$ and chemical potential $\mu$ is given by
\begin{equation}
Z=\rm{Tr} ~\rm{e}^{- \beta \left( H - \mu N \right)}, ~~~ 
\beta = (k_{B} T)^{-1} , 
\label{partition}
\end{equation}
where $k_B$ is the Boltzmann constant, $H$ is the Hamiltonian, 
and $N$ is particle number operator
\begin{equation}
N=\int d^3x ~ \psi^{\dagger}(x) \psi (x).
\label{PartNumber}
\end{equation}
The energy density of the system with free quarks is given by \cite{Li}
\begin{equation}
\epsilon = {1 \over V}{1 \over Z} \rm{Tr} ~ H ~ \rm{e}^{- \beta \left( H- \mu N \right)}
= -{1 \over V} \frac{\partial \ln Z} { \partial \beta } 
\vert_{\mu \beta} .
\end{equation} 
Going over to $T \to 0$, the energy density 
(where the contribution of $\mu=0$ is subtracted) becomes
\begin{equation}
\epsilon_{sub} =
{4 \pi \over (2 \pi)^{4}}
\int_{-\infty}^{\infty} d^3 p 
~\Theta \left( \mu - \sqrt{\sum_{j=1}^3 p_j^2 +m^2} \right)
\sqrt{\sum_{j=1}^3 p_j^2 +m^2}.
\end{equation}
Here $\Theta$ is the step function.
In the chiral limit $m\to 0$ one obtains
\begin{eqnarray*}
\epsilon_{sub} &=& {1 \over 4 \pi^3} \int_{-\infty}^{\infty} d^3 p
~ \Theta \left( \mu -\vert \vec{p} \vert \right) \vert \vec{p} \vert
\end{eqnarray*}
\begin{eqnarray}
& = &
{1 \over \pi^2} \int_0^{\mu} \vert \vec{p} \vert^3 d \vert \vec{p} \vert
={\mu^4 \over 4 \pi^2}.
\label{cont}
\end{eqnarray}
In the Hamiltonian formulation of LGT, Eq.(\ref{partition}) in well defined.
For Wilson fermions or Kogut-Susskind fermions,
the relation Eq.(\ref{cont}) is satisfied (see below).
However, if one constructs
the  fermionic lattice Lagrangian via Legendre transformation 
of the Hamiltonian, one cannot reproduce the continuum relation Eq.(\ref{cont}).  
Let us take the naive fermions as an example.
The action obtained via Legendre transformation of $H$ reads 
\begin{eqnarray}
S_f= a^4 \sum_x  m \bar{\psi}(x) \psi(x) 
+ {a^3 \over 2} \sum_{x}\sum_{k=\pm 1}^{\pm 4} 
\bar{\psi}(x) \gamma_k \psi(x+\hat{k})
+ a^4 \mu \sum_x \psi^{\dagger}(x) \psi (x),
\label{naive}
\end{eqnarray}
where $\gamma_{-k}=-\gamma_{k}$. 
This action gives the following result for the subtracted energy density 
\begin{equation}
\epsilon_{sub}={-a^{-4}  \over 4 \pi^{4}}
\int_{-\pi}^{\pi} d^4 p {\sum_{j=1}^3 \sin^2p_j +(ma)^2 \over
(\sin p_4 -i \mu a)^2 +\sum_{j=1}^3 \sin^2p_j +(ma)^2} -  \left[\mu=0 \right].
\label{latt}
\end{equation}
Taking the limit $m \to 0$ and the continuum limit $a \to 0$, 
$\epsilon_{sub} \propto (\mu / a)^2$, i.e. becoming quadratically divergent, 
and therefore it is inconsistent with the continuum result of Eq.(\ref{cont}).
This problem is not due to the species doubling of naive fermions, because
the case of Kogut-Susskind fermions or Wilson fermions is similar.

\bigskip

\noindent Hasenfratz and Karsch \cite{Hasenfratz} proposed the following solution:
If  $(\sin p_4-i\mu)^2$ is replaced by $\sin^2 (p_4-i\mu)$
the continuum result Eq.(\ref{cont}) is reproduced, 
except for a factor of 16. 
Correspondingly in the action, the chemical potential is introduced 
in the following way
\begin{eqnarray}
S_f &=& a^4 \sum_x  m \bar{\psi}(x) \psi(x) 
+ {a^3 \over 2} \sum_{x} \sum_{j=1}^3 
\left[ \bar{\psi}(x) \gamma_j \psi(x+\hat{j})
-\bar{\psi}(x+\hat{j}) \gamma_j \psi(x) \right] 
\nonumber \\
&+& {a^3 \over 2} \sum_x  
\left[ e^{\mu a} \bar{\psi}(x) \gamma_4 \psi (x+\hat{4})
-e^{-\mu a} \bar{\psi}(x+\hat{4}) \gamma_4 \psi (x)\right].
\label{Hansen}
\end{eqnarray}
The chemical potential can be introduced analogously for KS 
as well as for Wilson fermions.
Such treatment of the chemical potential is numerically feasible 
in the quenched approximation (where the fermionic determinant $\det \Delta$ 
is constraint to be $1$, and quark loops are suppressed).
However, there is evidence \cite{Quenched} that the quenched approximation
produces an unphysical onset of the critical chemical potential at the value
$\mu_{C}=M_{\pi} (m \not= 0)/2$, 
being in conflict with other theoretical predictions $\mu_C \approx M_N^{(0)}/3$ 
($M_N^{(0)}$ is the nucleon mass at $\mu=0$ and
$M_{\pi}(m \not=0)$ is the pion mass at finite bare quark mass $m$. 
A finite bare quark mass must be introduced 
in most of the numerical simulations). 
The unphysical onset of $\mu_C$ is considered as a defect of
the quenched approximation.

\bigskip

\noindent For full QCD, the fermionic degrees of freedom 
have to be integrated out. 
In the measure occurs the fermionic determinant $\det \Delta$.
For finite chemical potential $\det \Delta$ 
becomes complex (complex action problem), which renders numerical simulations
extremely difficult. 
Much effort has been made to solve the notorious complex
action problem: \\
\noindent (1) The Glasgow group has suggested to treat 
$\det \Delta$ as observable \cite{Glasgow}
This method requires a very large number of configurations, 
in particular for $\mu \approx \mu_C$. Even on a very small lattice $V=4^4$, 
the computational costs exceed the current computer capacity \cite{Zaragoza}. \\
\noindent (2) In the imaginary chemical potential 
method \cite{Imaginary} $\det \Delta$ becomes real, which works well  
for numerical simulations at high temperature and low density. 
But it might not work at low temperature and high density. \\
\noindent (3) It has been proposed to utilize a special symmetry \cite{Hands}. 
This is the only successful method in Lagrangian lattice QCD, 
but it works only for the SU(2) gauge group. \\
\noindent (4) Recently, a new approach has been proposed in \cite{Wiese}, 
using quantum spin variables. 
It remains to be seen whether this can be applied to QCD.\\

\section{Hamiltonian Approach}
\label{Our_Approach}
\subsection{Free fermions at zero chemical potential}
The lattice Hamiltonian describing noninteracting Wilson fermions 
in $d+1$ dimensions at $\mu=0$ reads
\begin{equation}
H= \sum_x  m \bar{\psi}(x) \psi(x) 
+ {1 \over 2a} \sum_{x,k=\pm j} \bar{\psi}(x) \gamma_k \psi(x+\hat{k}) 
+ {r \over 2a} \sum_{x,k=\pm j}  
\left[ \bar{\psi}(x)\psi (x)
-\bar{\psi}(x) \psi (x+\hat{k})\right].
\label{Hamiltonian_mu0}
\end{equation}
We want to diagonalize $H$ so that the 
fermionic field $\psi$ can be expressed in terms of 
up and down 2-spinors $\xi$ and $\eta^{\dagger}$,
\begin{equation}
\psi =\left( 
\begin{array}{c}
\xi \\
\eta^{\dagger} 
\end{array}
\right).
\label{updown}
\end{equation}
We define the bare vacuum state $\vert 0 >$ as 
\begin{equation}
\xi \vert 0\rangle = \eta \vert 0\rangle =0.
\end{equation}
Since the up and down components are coupled via the $\gamma_k$ matrices, 
the bare vacuum is not an eigenstate of $H$. 
Let $\vert \Omega >$ denote the physical vacuum state, 
and $E_{\Omega}$ the vacuum energy.
One can use a unitary transformation  
to decouple the up and down components \cite{Luo},
\begin{eqnarray}
H'= \exp(-iS) ~ H  ~ \exp(iS).
\end{eqnarray}
Such a transformation is similar to the 
Foldy-Wouthuysen transformation \cite{Bjorken}.
Then the physical vacuum state of $H$ can be expressed as
\begin{equation}
\vert \Omega \rangle = \exp(iS) \vert 0 \rangle . 
\end{equation}
The operator $S$ can be computed explicitly. 
For Wilson ($r\not= 0$) or naive ($r=0$) fermions it reads \cite{Luo}
\begin{eqnarray}
S &=& \sum_p \theta_p S_p,
\nonumber \\
S_{p} &=& -{1 \over A_{p}}
\sum_{j=1}^d \psi^{\dagger}_p \gamma_{j} \psi_p {\sin p_j a \over a},
\nonumber \\
A_{p} &=& \left[ \sum_{j=1}^d \left({\sin p_j a \over a}\right)^2 \right]^{1/2},
\label{operator}
\end{eqnarray}
and $p$ is the momentum. The transformed Hamiltonian becomes
\begin{eqnarray}
H'  &=&
\sum_{p} \left[ \left[ m+ 
{2r \over a} \sum_{j=1}^d \sin^2 \left( p_j a/2 \right) \right] \cos 2\theta_p
+ A_{p} \sin 2\theta_p \right] \bar{\psi}_p \psi_p 
\nonumber \\
&+& 
\left[ \cos 2\theta_p - \left[  m+ 
{2r \over a} \sum_{j=1}^d \sin^2 \left( p_j a/2 \right) \right] 
{\sin 2\theta_p \over A_p} \right] 
\sum_{j=1}^d \bar{\psi}_p i \gamma_j {\sin p_ja \over a}
\psi_p.
\label{H'}
\end{eqnarray}
The vacuum energy is given by
\begin{eqnarray}
E_{\Omega} &=&
\langle \Omega \vert H \vert \Omega \rangle
= \langle 0 \vert H' \vert 0 \rangle
\nonumber \\
&=& -2N_{c}N_{f}
\sum_{p} \left[ \left[ m+ 
{2r \over a} \sum_{j=1}^d \sin^2 \left( p_j a/2 \right) \right] \cos 2\theta_p
+ A_{p} \sin 2\theta_p \right],
\end{eqnarray}
where $N_c$ and $N_f$, respectively, are the number of colors 
and number of flavors. 
The vacuum energy $E_{\Omega}$ is minimized 
under variation of the parameters $\theta_p$ if 
\begin{equation}
\mbox{tan} ~ 2 \theta_{p} ={A_{p} \over 
m+ {2r \over a} \sum_{j=1}^d \sin^2 \left( p_j a/2 \right)}.
\label{theta}
\end{equation}
This condition also leads to the
cancellation of the second term in Eq. (\ref{H'})
coupling the up and down components such that 
\begin{equation}
H' \vert 0 \rangle =
\sum_{p} 
A'_p
\bar{\psi}_p \psi_p \vert 0 \rangle
= E_{\Omega}  \vert 0 \rangle,
\label{H'0}
\end{equation}
where we denote
\begin{equation}
A'_p=
\left[ \left[ m+ 
{2r \over a} \sum_{j=1}^d \sin^2 \left( p_j a/2 \right) \right]^{2}
+ A_{p}^{2} \right]^{1/2}. 
\end{equation}
The vacuum energy becomes
\begin{equation}
E_{\Omega}=-2N_{c}N_{f}\sum_{p} A'_p.
\end{equation}
It can be easily seen that $\vert \Omega \rangle$ is the eigenstate of $H$ and
$E_{\Omega}$ is its eigenvalue. For Wilson fermions, 
in the continuum limit $a \to 0$, for any finite momentum $p$, we have
\begin{equation}
A'_p \to \sqrt{m^2+p^2},
\label{dispersion}
\end{equation}
giving the correct dispersion relation.

\subsection{Free fermions at nonzero chemical potential}
\label{free_fermion_mu}
We follow the same steps as in the case $\mu=0$. 
According to Eq. (\ref{partition}), the role of the Hamiltonian is now played by
\begin{eqnarray}
H_{\mu}=H- \mu N,
\label{Hamiltonian}
\end{eqnarray}
where $H$ is given by Eq.(\ref{Hamiltonian_mu0}) and $N$ is given by 
Eq.(\ref{PartNumber}).
Let us define the state
$\vert n_p, \bar{n}_p  \rangle$ by
\begin{eqnarray}
&& \xi_p \vert 0_p, \bar{n}_p  \rangle=0,
~~~ \xi^{\dagger}_p \vert 0_p, \bar{n}_p  \rangle=\vert 1_p, 
\bar{n}_p  \rangle, ~~~ \xi_p \vert 1_p, 
\bar{n}_p  \rangle=\vert 0_p, \bar{n}_p  \rangle,
~~~ \xi^{\dagger}_p \vert 1_p, \bar{n}_p  \rangle=0, 
\nonumber \\
&& \eta_p \vert n_p, 0_p  \rangle=0,
~~~ \eta^{\dagger}_p \vert n_p, 0_p  \rangle=\vert n_p, 1_p  \rangle, 
~~~ \eta_p \vert n_p, 1_p  \rangle=\vert n_p, 0_p  \rangle,
~~~ \eta^{\dagger}_p \vert n_p, 1_p  \rangle=0 .
\label{state}
\end{eqnarray}
The numbers $n_p$ and $\bar{n}_p$ take the values 0 
or 1 due to the Pauli principle. 
By definition, the up and down components of the fermion field
are decoupled. Obviously, this is not an eigenstate of $H_{\mu}$ due to
the nondiagonal form of $H$.
However, they are eigenstates of $H'_{\mu}$, 
which are related to $H_{\mu}$ by a unitary transformation
\begin{equation}
H_{\mu}'= \exp(-iS) ~ H_{\mu}  ~ \exp(iS)=H'- \mu N.
\label{unitary_mu}
\end{equation}
For the vacuum eigenstate of $H_{\mu}$ we make an ansatz of the following form
\begin{equation}
\vert \Omega \rangle =  \exp(iS)\sum_p f_{n_p, \bar{n}_p} 
\vert n_p, \bar{n}_p  \rangle.
\end{equation}
$S$ is given by Eq.(\ref{operator}) and the parameter $\theta_p$
is given by Eq.(\ref{theta}). Both $S$ and $\theta_p$ do not depend on $\mu$ 
because the quark number operator $N$ commutes with $S$.
$H'$ is given by Eq. (\ref{H'}).
The vacuum energy thus obeys
\begin{eqnarray}
E_{\Omega} &=& \langle \Omega \vert H _{\mu} \vert \Omega \rangle
=\sum_{p',p} f_{n_{p'}, \bar{n}_{p'}} f_{n_p \bar{n}_p} 
\langle n_{p'} \bar{n}_{p'} \vert H_{\mu}'
\vert n_p, \bar{n}_p \rangle
\nonumber \\
&=& \sum_{p} C_{n_p, \bar{n}_p} 
\langle n_p, \bar{n}_p \vert H' -\mu N
\vert n_p, \bar{n}_p \rangle,
\end{eqnarray}
where we have introduced the notation $C_{n_p, \bar{n}_p}=f_{n_p, \bar{n}_p}^2$.
From Eq.(\ref{state}) follows 
\begin{eqnarray}
E_{\Omega} &=&
\sum_{p} C_{n_p, \bar{n}_p} \left( 
A'_p \langle n_p, \bar{n}_p \vert  \bar{\psi}_p \psi_p \vert n_p, \bar{n}_p \rangle
-\mu \langle n_p, \bar{n}_p \vert  {\psi}_p^{\dagger} 
\psi_p \vert n_p, \bar{n}_p \rangle
\right)
\nonumber \\
&=& 2N_c N_f 
\sum_{p} C_{n_p, \bar{n}_p} \left[ \left( A'_p-\mu \right) n_p  
+ \left( A'_p+ \mu \right)\bar{n}_p - A'_p- \mu \right].
\label{E_mu}
\end{eqnarray}
We have not yet specified the function $C_{n_p, \bar{n}_p}$.
For this purpose we use the condition of stability of the vacuum.   
Because $\mu >0$, the vacuum energy increases with $n_p$.
This means the vacuum is unstable unless $\bar{n}_p=0$.
This simplifies Eq. (\ref{E_mu}) to
\begin{equation}
E_{\Omega} =
2N_c N_f 
\sum_{p} C_{n_p} \left[ \left( A'_p-\mu \right) n_p  
- A'_p - \mu \right] ,
\end{equation}
where we use the abbreviation $C_{n_p}=C_{n_p,0}$. From the 
normalization condition $C_{0_p}+C_{1_p}=1$,
we obtain
\begin{equation}
E_{\Omega} =
2N_c N_f 
\sum_{p} \left[ C_{1_p}\left( A'_p-\mu \right)  
- A'_p - \mu \right].
\end{equation}
$C_{1_p}$ depends on the value of $\mu$ and its dependence can be seen 
by inspection of the derivative
\begin{equation}
{\partial E_{\Omega} \over \partial C_{1_p}}
= 2N_c N_f \left( A'_p-\mu \right).
\label{deriveritive_free}  
\end{equation}
For $\mu > A'_p$, the right-hand side 
is negative. Maximizing $C_{1_p}$ means minimizing the vacuum energy. 
Therefore, $C_{1_p}=1$.
On the other hand, for $\mu < A'_p$, the right-hand side 
is positive and for any $C_{1_p}$ the vacuum is unstable. 
Therefore, $C_{1_p}=0$. We can summarize these properties by writing
\begin{equation}
C_{1_p} = \Theta \left( \mu -A'_p\right).
\label{deriveritive}  
\end{equation}
Thus the vacuum energy becomes 
\begin{equation}
E_{\Omega} = 2N_c N_f 
\sum_{p} \left( C_{1_p} A'_p  - A'_p \right).
\end{equation}
The subtracted energy density reads
\begin{equation}
\epsilon_{sub} = \frac{ E_{\Omega} - E_{\Omega}\vert_{\mu=0} }
{ N_c N_f N_s }
= {2 \over N_s} 
\sum_{p} C_{1_p} A'_p
= {2\over \left(2 \pi \right)^3} \int_{-\infty}^{\infty} d^3p 
~ A'_p ~ \Theta \left(\mu -A'_p \right).
\end{equation}
Here $N_s$ is the number of spatial lattice sites.
In case of Wilson fermions, 
for $m=0$ and in the continuum $a= 0$,
for any finite momentum $p$, one has $A'_p = \vert p \vert$.
In 3+1 dimensions, at the corners of the Brillouin zone 
$p_j a = (\pi,0,0)$, $(0,\pi,0)$, 
$(0,0, \pi)$, $(\pi,\pi,0)$, $(0,\pi,\pi)$, $(\pi,0,\pi)$, $(\pi,\pi,\pi)$, 
one has 
$\Theta \left(\mu -A'_p \right)=0$.
Therefore, in the continuum we find
\begin{equation}
\epsilon_{sub}
={8 \pi \over \left(2 \pi \right)^3} \int_{0}^{\mu} ~ p ~ d^3p 
={\mu^4 \over 4 \pi^2}.
\label{lat_hamil}
\end{equation}
Thus we have proven that
we can reproduce in the Hamiltonian formulation 
the continuum result of the vacuum energy density, Eq. (\ref{cont}).
For naive fermions, in the continuum limit $a=0$, 
there will be an extra factor of $2^d$.

\subsection{Strong coupling QCD at nonzero chemical potential}
\subsubsection{Structure of the Hamiltonian}
As is well known, lattice QCD at $\mu=0$ confines
quarks and spontaneously breaks chiral symmetry.
For a sufficiently large chemical potential, this picture may change.
At lattice spacing $a \not= 0$, as discussed in Sect. \ref{Status},
none of the standard approaches to lattice fermions is satisfactory.
Here we set out to investigate finite density QCD 
in the strong coupling regime $1/g^2 << 1$, using the Hamiltonian formulation. 
One of the goals is to get a better understanding of the mechanism 
of chiral phase transition.
According to Ref. \cite{Luo},
$H'$ in Eq. (\ref{unitary_mu}) now is replaced by
\begin{eqnarray}
H' &=& \left[ m\left[1-\left( 2 \theta_0 \right)^2 d \right] 
+{\left( 2 \theta_0 \right)d \over a} \right]\sum_x \bar{\psi}(x) \psi(x) 
\nonumber \\
&+& {g^2 C_N \left( 2 \theta_0 \right)^2 \over 8aN_c} 
\sum_{x} \sum_{k=\pm j} \psi^{\dagger}_{c_1,f_1}(x) \gamma_k \psi_{c_2,f_1}(x+\hat{k})
\psi^{\dagger}_{c_2,f_2}(x+\hat{k}) \gamma_k \psi_{c_1,f_2}(x), 
\label{Hamiltonian_tran}
\end{eqnarray}
where $d=3$ denotes the spatial dimension, $c_1, ~c_2$ are color indices
and  $f_1, ~f_2$ are flavor indices (summation over repeated indices is understood), 
$\theta_0=1/(4ma+g^2C_N)$, and $C_N=(N_c^2-1)/(2N_c)$.
The four-fermion interaction is induced by gauge interactions with fermions.
A very similar Hamiltonian has been derived in Ref. \cite{SG}
using strong coupling and large $N_c$ expansion. 
After a Fierz transformation, $H'$ becomes \cite{Luo}
\begin{eqnarray}
H' &=& \left[ m\left[ 1-\left( 2 \theta_0 \right)^2 d \right] 
+{\left( 2 \theta_0 \right)d \over a} \right] \sum_x \bar{\psi}(x) \psi(x)
+ {g^2 C_N d \left( 2 \theta_0 \right)^2 \over 4a}\sum_x \psi^{\dagger}(x) \psi(x) 
\nonumber \\
&-& {g^2 C_N \left( 2 \theta_0 \right)^2 \over 32aN_c} 
\sum_{x} \sum_{k=\pm j} L_A \psi^{\dagger}_{f_1}(x) \Gamma_A \psi_{f_2}(x)
\psi^{\dagger}_{f_2}(x+k) \Gamma_A \psi_{f_1}(x+k) .
\label{Hamiltonian_fierz}
\end{eqnarray}
The matrices $\Gamma_A$ and $L_A$ are given in Tab.[1].

\begin{table}
\begin{center}
\begin{tabular}{|c|c|c|c|c|c|c|c|c|}
\hline
$\Gamma_A$ & 1 & $\gamma_j$ & $\gamma_4$ & $\gamma_5$ 
& $i\gamma_4 \gamma_5$ & $i\gamma_4 \gamma_j$ &
$i \epsilon_{j j_1 j_2} \gamma_{j_1} \gamma_{j_2}$ &
$i \epsilon_{j j_1 j_2} \gamma_4 \gamma_{j_1} \gamma_{j_2}$\\
\hline
$L_A$ & 1 & $-1+2\delta_{k,j}$ & -1 & -1 & 1 & $1-2\delta_{k,j}$ &
$-1+2\delta_{k,j}$ &
$1-2\delta_{k,j}$\\
\hline
\end{tabular}
\end{center}
\caption{$\Gamma$ matrices and coefficients.}
\end{table}

\noindent Let us define the following operators \cite{Luo,Guo},
\begin{eqnarray}
&& \Pi(x)_{f_1 f_2}={1 \over 2 \sqrt {- \bar{v} }}
\psi^{\dagger}_{f_1}(x)\left(1-\gamma_4 \right) \gamma_5  \psi_{f_2}(x), 
\nonumber \\
&& V_j(x)_{f_1 f_2}={1 \over 2 \sqrt {- \bar{v}}}
\psi^{\dagger}_{f_1}(x)\left(1-\gamma_4 \right)\gamma_j \psi_{f_2}(x), 
\label{mesons}
\end{eqnarray}
where
\begin{equation}
\bar{v}=
{1\over N_f N_s}\sum_{p} C_{n_p, \bar{n}_p} 
\langle n_p, \bar{n}_p \vert  \bar{\psi}_p \psi_p \vert n_p, \bar{n}_p \rangle
=
{2N_c \over N_s}
\sum_{p} C_{n_p, \bar{n}_p} \left( n_p + \bar{n}_p -1 \right).
\end{equation}
Using mean field approximation, one can show that \cite{Guo}
\begin{eqnarray}
&& \left[ \Pi(x), \Pi^{\dagger}(x') \right] =\delta_{x,x'},
\nonumber \\
&& \left[ V_j(x), V_j^{\dagger}(x') \right] =\delta_{x,x'}
\label{mesons_com}
\end{eqnarray}
Thus the operators $\Pi$ and $V_{j}$, defined in Eq. (\ref{mesons}), behave like 
pseudoscalar and vector operators. In Ref.\cite{Guo} it has been shown
that the operator $\bar{\psi}\psi$ satisfies the same commutation relations
as $\bar{v}+2\Pi^{\dagger}\Pi+2\sum_j V^{\dagger}_j V_j$. 
%Without loss of generality, we will consider the $N_f=1$ case. 
Therefore, $H'$
in Eq. (\ref{Hamiltonian_fierz}) can be
written in terms of pseudo-scalar and vector particle operators 
in the following way
\begin{eqnarray}
H' &=& E_{\Omega}^{(0)}+G_1 \sum_x 
\left( \Pi^{\dagger}(x) \Pi(x) + \sum_j V_j^{\dagger}(x) V_j(x) \right)
\nonumber \\
&+& {G_2 \over 2}
\sum_{x,k}\left( \Pi^{\dagger}(x) \Pi^{\dagger}(x+k) 
+ \sum_j V_j^{\dagger}(x) V_j^{\dagger}(x+k) \left(1-2\delta_{jk} \right)
+h.c. \right),
\label{Hamiltonian_mesons}
\end{eqnarray}
where
\begin{eqnarray}
E_{\Omega}^{(0)} &=& N_f N_s
\left[ m\left[1-\left( 2 \theta_0 \right)^2 d \right] 
+{\left( 2 \theta_0 \right)d \over a} \right]  \bar{v} 
\nonumber \\
&+& N_f N_s {g^2 C_N d \left( 2 \theta_0 \right)^2 \over 4a} v^{\dagger} 
- N_f N_s {g^2 C_N \left( 2 \theta_0 \right)^2 d  \over 16aN_c}  
\left( v_2^{\dagger}- \bar{v}_2 \right),
\nonumber \\
G_1 &=&
2 \left[ m\left[1-\left( 2 \theta_0 \right)^2 d \right] 
+{\left( 2 \theta_0 \right)d \over a} \right] 
+ {g^2 C_N d \left( 2 \theta_0 \right)^2 \over 4 a N_c} \bar{v},
\nonumber \\
G_2 &=& - {g^2 C_N \left( 2 \theta_0 \right)^2 \over 8 a N_c}  \bar{v},
\nonumber \\
v^{\dagger} &=&
{1\over N_f N_s}\sum_{p} C_{n_p, \bar{n}_p} 
\langle n_p, \bar{n}_p \vert  \psi_p^{\dagger} \psi_p \vert n_p, \bar{n}_p \rangle
= {2N_c \over N_s}
\sum_{p} C_{n_p, \bar{n}_p} \left( n_p - \bar{n}_p + 1 \right),
\nonumber \\
v_2^{\dagger} &=&
{1\over N_f N_s}\sum_{p} C_{n_p, \bar{n}_p} 
\langle n_p, \bar{n}_p \vert  \psi_{f_1,p}^{\dagger} \psi_{f_2,p} 
\psi_{f_2,p}^{\dagger} \psi_{f_1,p} \vert n_p, \bar{n}_p \rangle
\nonumber \\
&=& {(2N_c)^2  \over N_s}
\sum_{p} C_{n_p, \bar{n}_p} \left( n_p - \bar{n}_p + 1 \right)^2,
\nonumber \\
\bar{v}_2 &=&
{1\over N_f N_s}\sum_{p} C_{n_p, \bar{n}_p} 
\langle n_p, \bar{n}_p \vert  \bar{\psi}_{f_1,p} \psi_{f_2,p} 
\bar{\psi}_{f_2,p} \psi_{f_1,p} 
\vert n_p, \bar{n}_p \rangle
\nonumber \\
&=&
{(2N_c)^2  \over N_s}
\sum_{p} C_{n_p, \bar{n}_p} \left( n_p + \bar{n}_p -1 \right)^2.
\label{Paramters}
\end{eqnarray}
In Eq.(\ref{Hamiltonian_mesons}),
we have ignored the nonmeson terms which give no contribution to the
energy. Making a Fourier transformation, one obtains
\begin{eqnarray}
H' &=& E_{\Omega}^{(0)}+G_1 \sum_p 
\left( \Pi^{\dagger}(p) \Pi(p) + \sum_j V_j^{\dagger}(p) V_j(p) \right)
\nonumber \\
&+& G_2 \sum_{p}\left( \Pi^{\dagger}(p) \Pi^{\dagger}(-p) +h.c. \right) 
\sum_j \cos p_j a 
\nonumber \\
&+& G_2 \sum_{p,j} \left( V_j^{\dagger}(p) V_j^{\dagger}(-p) + h.c. \right)  
\left(\sum_{j'} \cos p_{j'} a - 2 \cos p_j a \right) .
\label{Hamiltonian_mesons_for}
\end{eqnarray}
This can be diagonalized by a Bogoliubov transformation \cite{Luo}
\begin{eqnarray}
P(p) &=& \cosh u(p) a(p) + \sinh u(p) a^{\dagger}(-p),
\nonumber \\
V_j(p) &=& \cosh v_j(p) b(p) + \sinh v_j(p) b^{\dagger}(-p),
\label{Bogoliubov}
\end{eqnarray}
where
\begin{eqnarray}
\tanh  ~2u(p) &=& {-2G_2 \over G_1}\sum_j \cos p_j a,
\nonumber \\
\tanh  ~2v_j(p) &=& {-2G_2 \over G_1}\left(\sum_{j'} \cos p_{j'} a
-2 \cos p_j a \right) .
\label{Bogoliubov_con}
\end{eqnarray}
This condition also minimizes the vacuum energy.
The Bogoliubov transformed Hamiltonian eventually becomes
\begin{eqnarray}
H'' & = & E_{\Omega}^{(0)}
+ {N_f^2 \over 2} G_1 \sum_p \left[ \sqrt{1-\tanh^2 2u(p)} -1 \right]
+ {N_f^2 \over 2} G_1 \sum_{p,j} \left[ \sqrt{1-\tanh^2 2v_j(p)} -1 \right]
\nonumber \\
&+& G_1 \sum_p \sqrt{1-\tanh^2 2u(p)} a^{\dagger}(p) a(p)
+ G_1 \sum_{p,j} \sqrt{1-\tanh^2 2v_j(p)} b_j^{\dagger}(p) b_j(p) .
\label{Bogoliubov_H}
\end{eqnarray}

\subsubsection{Vacuum energy}
The vacuum energy is given by
\begin{eqnarray}
E_{\Omega} & = & \langle \Omega \vert H _{\mu} \vert \Omega \rangle
= E_{\Omega}^{(0)}
-  N_f N_s \mu v^{\dagger}
\nonumber \\
& + & {N_f^2 \over 2} G_1 \sum_p \left[ \sqrt{1-\tanh^2 2u(p)} -1 \right]
+ {N_f^2 \over 2} G_1 \sum_{p,j} \left[ \sqrt{1-\tanh^2 2v_j(p)}-1 \right].
\end{eqnarray}
From Eqs.(\ref{Paramters})-(\ref{Bogoliubov_H}), we get for $m=0$
\begin{eqnarray}
\frac{ E_{\Omega} }{ 2N_f N_c} &=&
\sum_{n_p,\bar{n}_p} C_{n_p,\bar{n}_p} \left[
\left( m_{dyn}^{(0)}- \mu \right) n_p
+\left(m_{dyn}^{(0)} + \mu + 2m_{dyn}^{(0)} n_p \right) \bar{n}_p \right]
\nonumber \\
&-& \left( \mu +m_{dyn}^{(0)} \right)
\sum_{n_p,\bar{n}_p} C_{n_p,\bar{n}_p} 
\nonumber \\
&+&  {N_f \over 2} G_1 \sum_p \left[ \sqrt{1-\tanh^2 2u(p)}-1 \right]
+ {N_f \over 2} G_1 \sum_{p,j} \left[ \sqrt{1-\tanh^2 2v_j(p)}-1 \right].
\end{eqnarray}
Here $m_{dyn}^{(0)}=d / (ag^2C_N)$ is the dynamical quark mass at $\mu=0$.
It is obvious that $\bar{n}_p=0$, otherwise, the vacuum is unstable.
Using the notation and normalization condition for the coefficient as in
Sect. \ref{free_fermion_mu}, we obtain
\begin{eqnarray}
{ E_{\Omega} \over 2N_c N_f} & = &
\sum_{p} C_{1_p} \left( m_{dyn}^{(0)}-\mu \right)
-\sum_{p} \left( m_{dyn}^{(0)}+\mu \right) 
\nonumber \\
&+&  {N_f \over 2} G_1 \sum_p \left[ \sqrt{1-\tanh^2 2u(p)}-1 \right]
+ {N_f \over 2} G_1 \sum_{p,j} \left[ \sqrt{1-\tanh^2 2v_j(p)}-1 \right]. 
\label{Energy_vacuum}
\end{eqnarray}
Again, using the same argument as in Sect. \ref{free_fermion_mu},
the coefficient $C_{1_p}$ must be
\begin{equation}
C_{1_p} = \Theta \left( \mu -m_{dyn}^{(0)}\right).
\end{equation}
Substituting into  Eq.(\ref{Energy_vacuum}) yields
\begin{eqnarray}
\frac{ E_{\Omega} }{ 2 N_c N_f N_s} & =&
\left(m_{dyn}^{(0)}-\mu \right) \Theta \left(\mu -m_{dyn}^{(0)}\right)
-m_{dyn}^{(0)}-\mu 
\nonumber \\
&+&  {N_f \over 2 N_s} G_1 \sum_p \left[ \sqrt{1-\tanh^2 2u(p)} -1 \right]
+ {N_f \over 2 N_s} G_1 \sum_{p,j} \left[ \sqrt{1-\tanh^2 2v_j(p)}-1 \right].
\label{result1} 
\end{eqnarray}

\subsubsection{Chiral condensate and critical $\mu$}
According to the Feynman-Hellmann theorem,
the chiral condensate is related to the ground state energy by
\begin{eqnarray}
\langle \bar{\psi} \psi \rangle =
{1 \over N_f N_s} \lim_{m\to 0} 
{\partial E_{\Omega} \left( m\not=0 \right)
\over \partial m}
= \langle \bar{\psi} \psi \rangle^{(0)}
\left[ 1-\Theta \left( \mu- m_{dyn}^{(0)}\right) \right] ,
\label{result2} 
\end{eqnarray}
where $\langle \bar{\psi} \psi \rangle^{(0)}$
is the chiral
condensate at $\mu=0$
\begin{eqnarray}
\langle \bar{\psi} \psi \rangle^{(0)}
=-2N_C \left( 1 - {4d \over g^4 C_N^2} \right) \left(1-{N_f \over N_c} I_1
-{N_f \over N_c} I_2 \right) 
\end{eqnarray}
and for $d=3$
\begin{eqnarray*}
I_1 & =& {1 \over 2(2\pi)^3} \int_{-\pi}^{\pi}  ~ d^3p^{\prime}
\left( {1 \over \sqrt{1-\left({1 \over 3} 
\sum_j \cos p_j^{\prime} \right)^2}} - 1\right)=0.078354 \pm 2 \times 10^{-6},
\end{eqnarray*}
\begin{eqnarray*}
I_2 & = & {1 \over 2(2\pi)^3} \sum_j\int_{-\pi}^{\pi}  ~ d^3p^{\prime}
\left( {1 \over \sqrt{1-\left({1 \over 3} 
\left( \sum_{j^{\prime}} \cos p_{j^{\prime}}^{\prime}  
- 2\cos p_j^{\prime}\right) \right)^2}} 
- 1\right)
\end{eqnarray*}
\begin{eqnarray}
& =  & 0.235075 \pm 4 \times 10^{-6}.
\end{eqnarray}

According to Eq. (\ref{result2}), for $\mu < m_{dyn}^{(0)}$, 
$\langle \bar{\psi} \psi \rangle
=\langle \bar{\psi} \psi \rangle^{(0)} \not=0$, i.e.,
chiral symmetry is spontaneously broken.
For $\mu > m_{dyn}^{(0)}$, $\langle \bar{\psi} \psi \rangle=0$,
i.e., chiral symmetry is restored. Therefore, there is a first order
chiral phase transition and the critical value of $\mu$ is given by
\begin{equation}
\mu_C = m_{dyn}^{(0)} = {d \over g^2 C_N a},
\label{result3} 
\end{equation}
The critical chemical potential $\mu_C$ is equal 
to the dynamical quark mass at $\mu=0$, 
which agrees with the result from an entirely different method \cite{Yaouanc88_1}.
(The authors argued this was a second order
phase transition, in contrast we clearly observe a first order transition).
Our result is consistent with other theoretical predictions 
$\mu_C \approx M_N^{(0)}/3$, 
because (see below) at $\mu=0$ holds $M_N^{(0)} \approx 3m_{dyn}^{(0)}$.

\subsubsection{Quark number density and susceptibility}
\label{number_density}
We can compute now the quark number density in the chiral limit $m=0$, which yields
\begin{equation}
n_q = {-1 \over 2 N_c N_f N_s}
{\partial E_{\Omega} 
\over \partial \mu}-1
=
\frac{ \langle \Omega \vert \sum_{x} \psi^{\dagger}(x) \psi (x) \vert \Omega \rangle } 
{2 N_c N_f N_s }-1
= \Theta \left(\mu-\mu_C \right),
\label{result4} 
\end{equation}
which is consistent with the $\beta=0 $ simulation results 
described in \cite{FirstOrder}, 
and however,  is different from 
the large $\mu$ behavior  in the continuum ({\it i.e.} the Stefann-Boltzmann law $n_q \propto \mu^3$).
It remains to be seen whether higher order $1/g^2$ calculations
will improve this behavior.

The quark number susceptibility, standing for the response of the 
quark number density to infinitesimal changes in $\mu$, is
\begin{equation}
\chi_q = 
\frac{ \partial n_q} 
{\partial \mu }
= \delta \left(\mu-\mu_C \right).
\label{result5} 
\end{equation}

\subsubsection{Mass spectrum}
Finally, let us look at some implications on the thermal mass spectrum
of the pseudoscalar meson, vector meson and nucleon.
The thermal mass is defined by
$M_{h}^{\star} = \langle h \vert H- \mu N \vert h \rangle -E_{\Omega}$.
For the pseudoscalar meson, in the chiral limit $m=0$,
\begin{eqnarray}
M_{\pi}^{\star}
 &= &G_1 \sqrt{1-\tanh^2 2u(p=0)}
=\left\{
\begin{array}{cc}
0 & ~~~ \rm{for} ~~~ \mu < \mu_C, \\
4m^{(0)}_{dyn} &~~~ \rm{for} ~~~ \mu > \mu_C . 
\end{array}
\right .
\label{result6} 
\end{eqnarray}
Therefore, in the broken phase,
the pseudoscalar is a Goldstone boson ($M_{\pi}^{\star} \propto \sqrt{m} \to 0$), 
and in the symmetric phase,
it is no longer a Goldstone boson. For the vector meson,
\begin{eqnarray}
M_V^{\star}
 &= &G_1 \sqrt{1-\tanh^2 2v_j(p=0)}
=\left\{
\begin{array}{cc}
M_V^{(0)} & ~~~ \rm{for} ~~~ \mu < \mu_C, \\
4m^{(0)}_{dyn} &~~~ \rm{for} ~~~ \mu > \mu_C . 
\end{array}
\right .
\label{result7} 
\end{eqnarray}
where $M_V^{(0)}=4 \sqrt{d-1}/(ag^2C_N)$ is the vector mass
at $\mu=0$. Therefore, $\partial M/\partial \mu 
\propto \delta(\mu-\mu_C)$ for the pseudoscalar and vector mesons. 
It is worth mentioning in Ref. \cite{MV}, the authors found 
$\partial M/\partial \mu = 0$ outside the critical region.
To see the critical behavior at zero temperature, 
one should be very close to $\mu_C$.
This behavior is consistent with that of the quark number density 
discussed in Sect. \ref{number_density}.
To see whether the meson thermal masses depend on $\mu$, 
higher order $1/g^2$ corrections must be included.

For the nucleon, we obtain the expected behavior
\begin{eqnarray}
M_N^{\star}
 &= & 
M_N^{(0)}- 3\mu
\label{result8}
\end{eqnarray}
for  $\mu < \mu_C$,
where $M_N^{(0)} \approx 3m_{dyn}^{(0)}$.
This leads to 
$M_N^{\star}=0$ at $\mu = \mu_C$.

\section{Outlook}
In this paper, we have developed a Hamiltonian approach 
to lattice QCD at finite density. 
It avoids the usual problem of either an incorrect continuum limit or a premature onset of the
transition to nonzero quark density as $\mu$ is raised.
The main result in the free case
is given by Eq. (\ref{lat_hamil}), and those in the strong coupling
regime are given by Eqs. (\ref{result1})-(\ref{result8}).
We have seen that the approach
works well in the free case and also in the strong coupling regime. 
We predict that at strong coupling, the chiral transition is of first order,
and the critical chemical potential $\mu_C \approx M_N^{(0)}/3$.

Here we have only considered zero temperature. In the case of finite 
temperature, contributions from thermal excitations 
will make the calculations quite complicated. 
We plan to address this issue in a future paper.

We are also aware that the strong coupling limit is not compatible 
with the continuum limit where $a \to 0$ and $1/g^2 \to \infty$. 
For pure gauge theory, within a Hamiltonian approach,
we can extend to the intermediate coupling 
and obtain meaningful results for the glueballs \cite{glueball}. 
For fermions, the calculation is far from trivial.
Recently we proposed a Monte Carlo technique 
in the Hamiltonian formulation \cite{mch} for the purpose to do 
nonperturbative numerical simulations, by combining the virtues of 
the Monte Carlo algorithm with importance sampling and the Hamiltonian approach.
We hope to apply it to QCD and with the aim 
to obtain useful information for RHIC physics.

\bigskip

\noindent
{\bf Acknowledgments}

X.Q.L. is grateful to V. Azcoiti, M.P. Lombardo, and S. Hands for useful discussions.
We also thank Z.H. Mei for participation at the initial stage of the project.
H.K. has been supported by NSERC Canada.
X.Q.L. is supported by the
National Science Fund for Distinguished Young Scholars (19825117),
National Natural Science Foundation (19605009, 19677205),
Hong Kong Foundation of the Zhongshan University 
Advanced Research Center (98P1).
S.H.G. and X.Q.L. are supported by the Ministry of Education,
the Doctoral Program of Higher Education and   
Guangdong Provincial Natural Science Foundation (990212) of China.
X.Q.L. and E.B.G. are supported by 
the Guangdong National Communication Development Ltd.

\end{document}